\documentclass{article}
\usepackage[utf8]{inputenc}
\usepackage{graphicx}
\usepackage{amsmath}
\usepackage{amssymb}
\usepackage[T1]{fontenc}
\usepackage{authblk}
\usepackage{xcolor}
\usepackage{soul}

\usepackage{cite}

\newcommand{\lp}{\ensuremath{\left(}}
\newcommand{\rp}{\ensuremath{\right)}}
\newcommand{\w}{\ensuremath{\omega}}

\newcommand{\ppz}{\ensuremath{{\partial \over \partial z}}}
\newcommand{\ppr}{\ensuremath{{\partial \over \partial r}}}
\newcommand{\e}{\ensuremath{\varepsilon}}

\newcommand{\rhat}{\ensuremath{\mathbf{\hat{r}}}}

\newcommand{\xhat}{\ensuremath{\mathbf{\hat{x}}}}
\newcommand{\yhat}{\ensuremath{\mathbf{\hat{y}}}}

\title{Computational inverse design\\for ultra-compact single-piece metalenses\\free of chromatic and angular aberration}
\author[1]{Zin~Lin\thanks{zinlin@mit.edu}}
\author[2]{Charles~Roques-Carmes}
\author[3,4]{Rasmus~E.~Christiansen}
\author[2,5]{Marin~Solja\v{c}i\'{c}}
\author[1]{Steven~G.~Johnson}
\affil[1]{Department of Mathematics, Massachusetts Institute of Technology, Cambridge MA 02138, USA}
\affil[2]{Research Lab of Electronics, Massachusetts Institute of Technology, Cambridge MA 02138, USA}
\affil[3]{Department of Mechanical Engineering, Technical University of Denmark, Nils Koppels All\'{e}, Building 404, 2800 Kongens Lyngby, Denmark}
\affil[4]{NanoPhoton---Center for Nanophotonics, Technical University of Denmark, {\O}rsteds Plads 345A, DK-2800 Kgs. Lyngby, Denmark}
\affil[5]{Department of Physics, Massachusetts Institute of Technology, Cambridge MA 02138, USA}

\date{\today}

\begin{document}

\maketitle

\begin{abstract}
We present full-Maxwell topology-optimization design of a single-piece multlayer metalens, about 10 wavelengths~$\lambda$ in thickness, that simultaneously focuses over a $60^\circ$ angular range and a 23\% spectral bandwidth without suffering chromatic or angular aberration, a ``plan-achromat.''  At all angles and frequencies it achieves diffraction-limited focusing (Strehl ratio $> 0.8$) and absolute focusing efficiency $> 50$\%.  Both 2D and 3D axi-symmetric designs are presented, optimized over $\sim 10^5$ degrees of freedom.  We also demonstrate shortening the lens-to-sensor distance while producing the same image as for a longer ``virtual'' focal length and maintaining plan-achromaticity.  These proof-of-concept designs demonstrate the ultra-compact multi-functionality that can be achieved by exploiting the full wave physics of subwavelength designs, and motivate future work on design and fabrication of multi-layer meta-optics.
  
\end{abstract}

\section{Introduction}

\begin{figure}[htb!]
    \centering
    \includegraphics[scale=0.3]{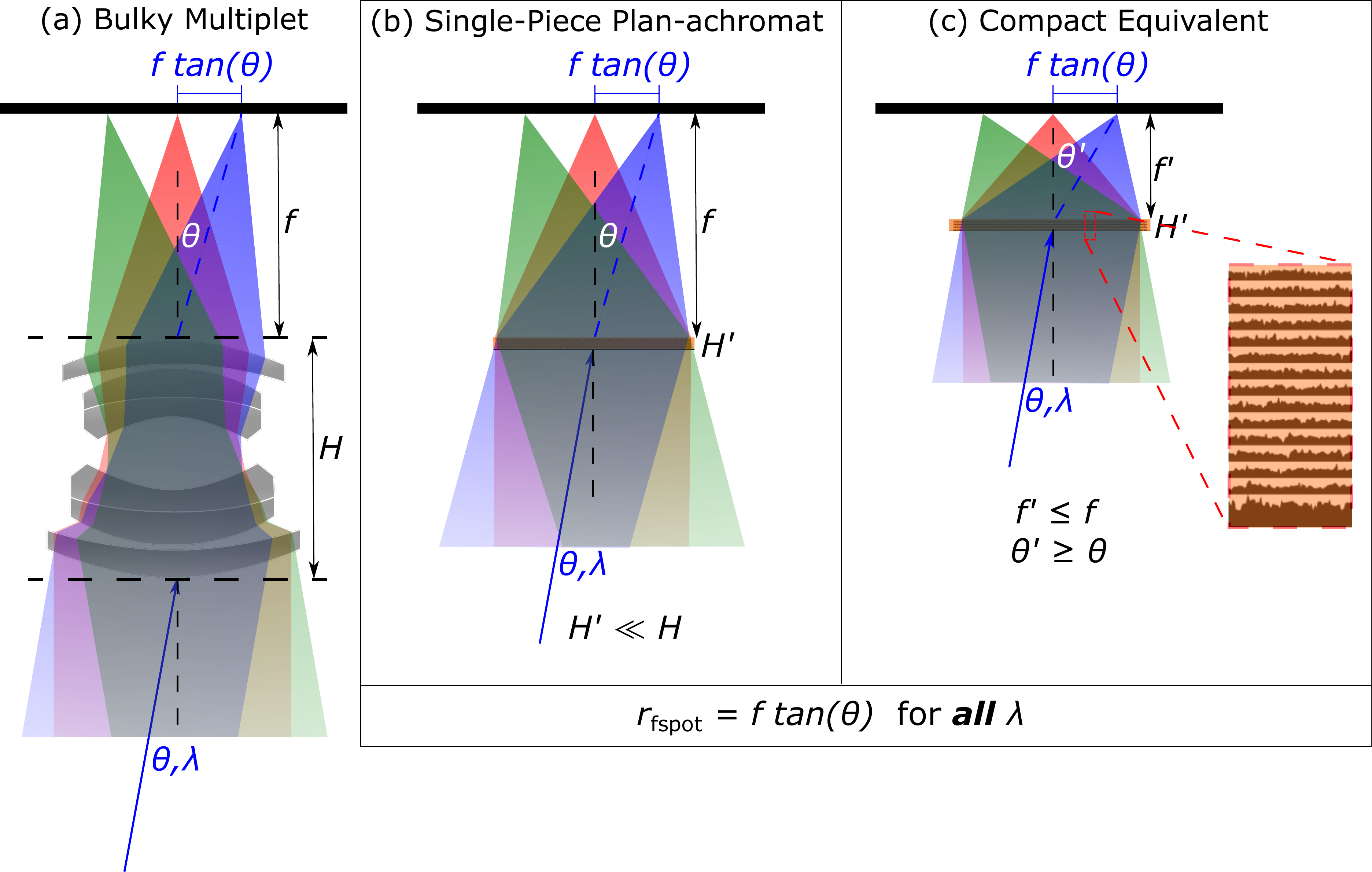}
    \caption{(a) A bulky multiplet (lens array) is needed to correct chromatic and geometric aberrations. (b) An ultra-compact nanophotonic \emph{single-piece} plan-achromat inverse-designed to achieve both chromatic and geometric aberration corrections. (c) Further miniaturization by designing a metastructure that can emulate a virtual focal distance $f < f'$ by stretching the out-going angles $\theta' > \theta$ while correcting aberrations.}
    \label{fig:scheme}
\end{figure}

\begin{figure}[ht!]
    \centering
    \includegraphics[scale=0.57]{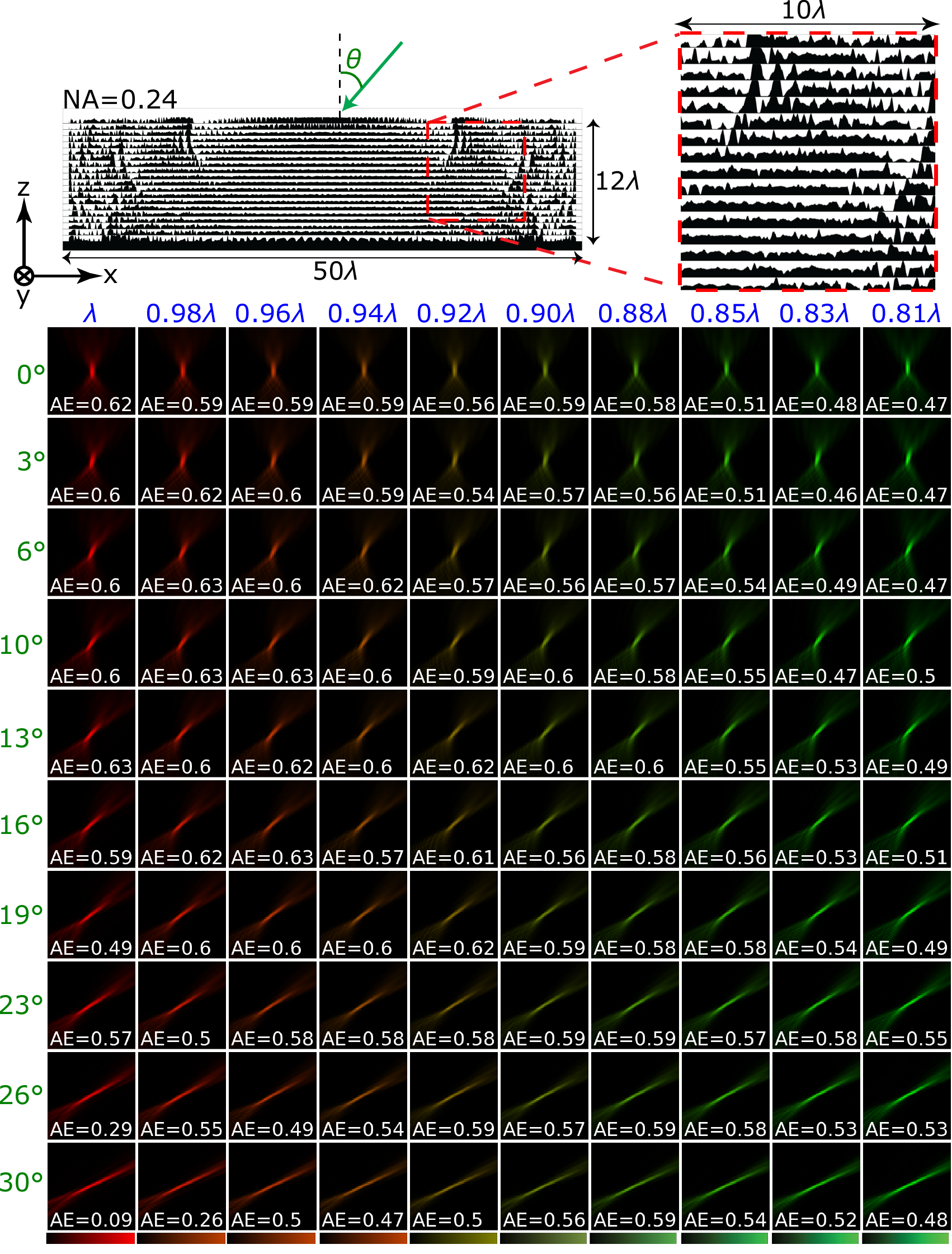}
    \caption{A 2D cylindrical metalens (invariant along $y$) that corrects both chromatic and geometric aberrations within a spectral bandwidth of 23\% and an angular bandwidth of 60$^\circ$. The lens is $50\lambda$ wide and $12\lambda$ thick (note the scale bar) with a numerical aperture (NA) of 0.24; it is made up of 20 layers of 3D-printable polymer (refractive index $= 1.5$). The lens achieves a Strehl ratio (SR) above 0.8 (diffraction limited focusing) for each of the design frequencies and angles while exhibiting an averaged absolute focusing efficiency (AE) of 55\%. }
    \label{fig:2d1}
\end{figure}

\begin{figure}[htb!]
    \centering
    \includegraphics[scale=0.4]{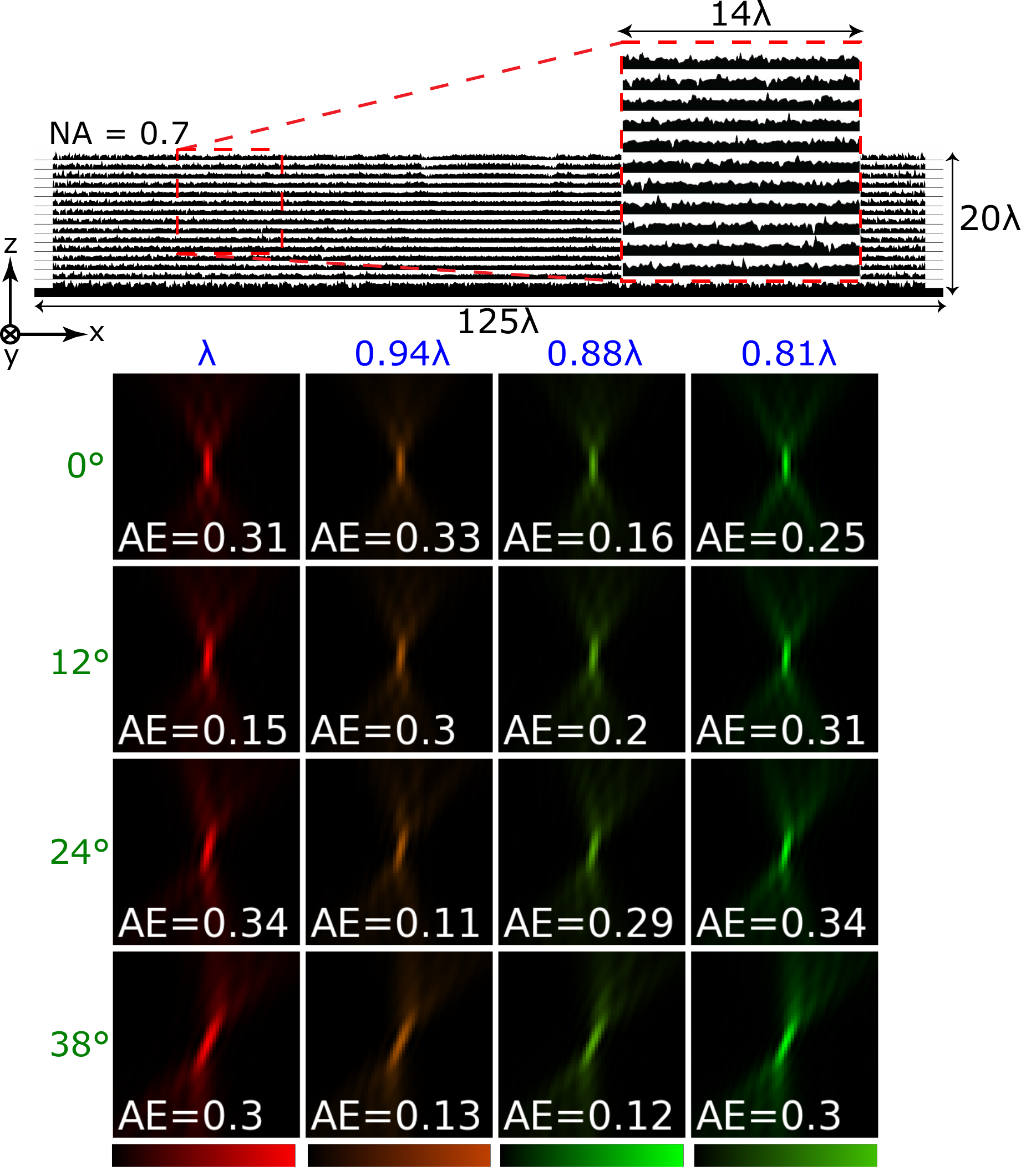}
    \caption{A high-NA 2D cylindrical metalens that simultaneously corrects chromatic and geometric aberrations at 4 wavelengths and 4 angles within a spectral bandwidth of 23\% and an angular bandwidth of 80$^\circ$. The lens is $125\lambda$ wide and $24\lambda$ thick (note the scale bar) with an NA of 0.7; it is made up of 15 layers of 3D-printable polymer (refractive index $= 1.5$). The lens achieves an SR above 0.8 (diffraction limited focusing) for each of the design frequencies and angles while exhibiting an averaged absolute efficiency of 25\%. }
    \label{fig:2d2}
\end{figure}

\begin{figure}[htb!]
    \centering
    \includegraphics[scale=0.55]{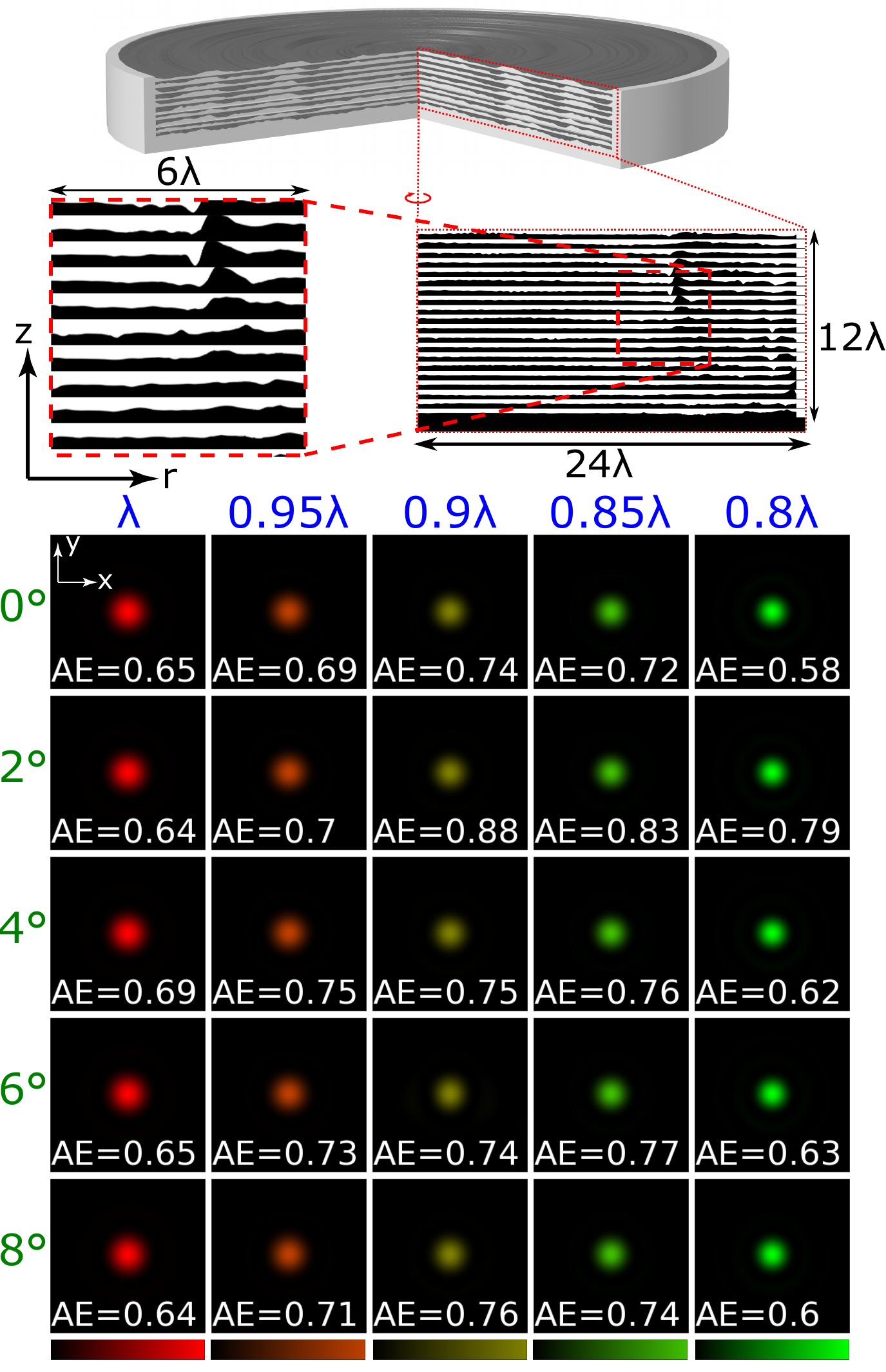}
    \caption{An axisymmetric 3D metalens that simultaneously corrects chromatic and geometric aberrations at 5 wavelengths and 5 angles within a spectral bandwidth of 23\% and an angular bandwidth of 16$^\circ$. The lens has a diameter of $50\lambda$ wide and a thickness of $12\lambda$ (note the scale bar) with an NA of 0.12; it is made up of 20 layers of 3D-printable polymer (refractive index $= 1.5$). The lens achieves an SR above 0.8 (diffraction limited focusing) for each of the design frequencies and angles. Averaged absolute focusing efficiency (AE) is 71\%.}
    \label{fig:3d}
\end{figure}

\begin{figure}[htb!]
    \centering
    \includegraphics[scale=0.55]{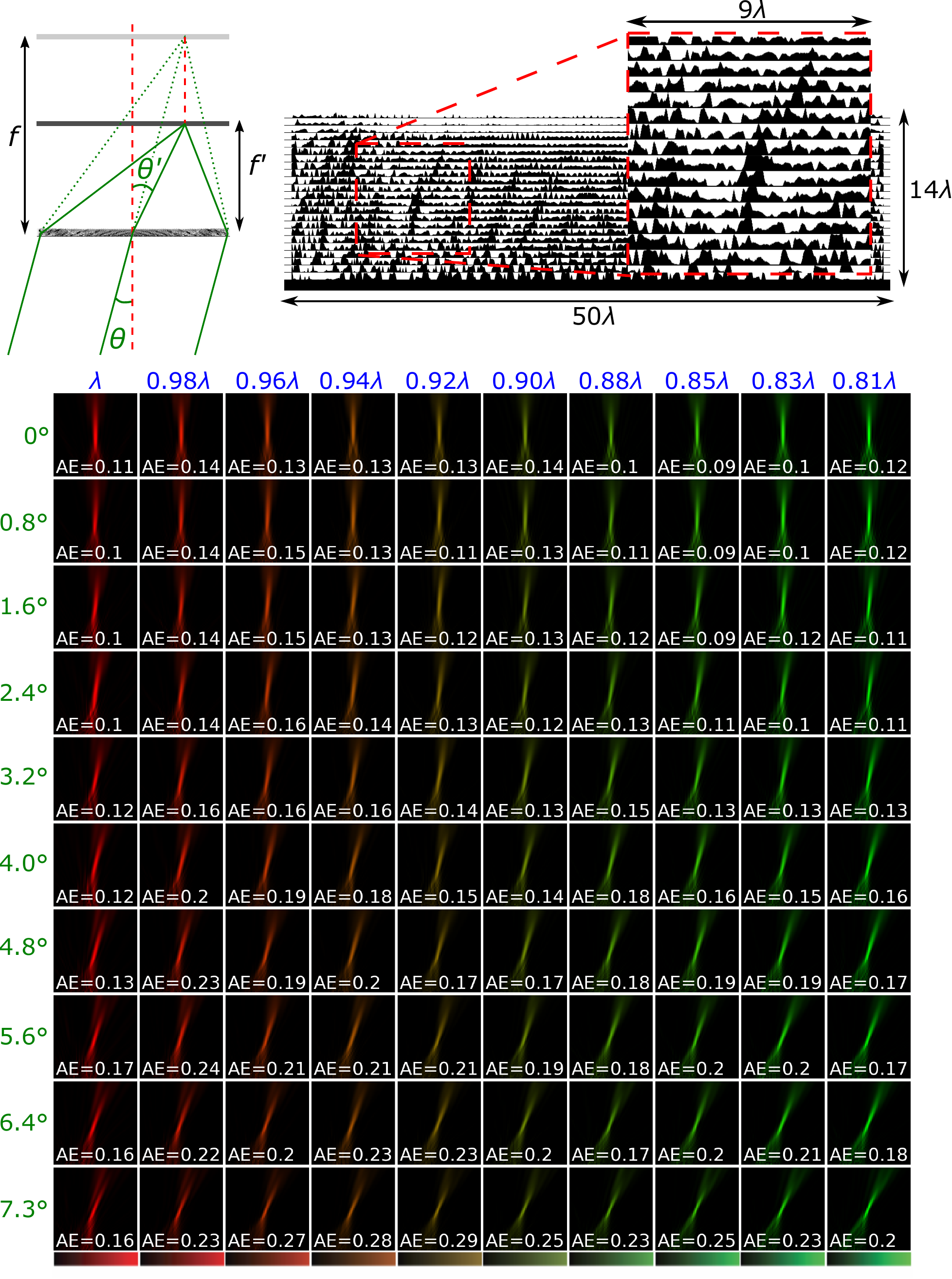}
    \caption{A 2D cylindrical metalens that emulates a virtual focal length of $200\lambda$ (54\% compression of lens-to-sensor distance) while simultaneously correcting chromatic and geometric aberrations at 10 wavelengths and 10 angles within a spectral bandwidth of 23\% and an angular bandwidth of 16$^\circ$. The lens has a diameter of $50\lambda$ wide and a thickness of $12\lambda$ thick (note the scale bar) with a \emph{virtual} NA of 0.12; it is made up of 23 layers of 3D-printable polymer (refractive index $= 1.5$). The lens achieves an SR above 0.8 (diffraction limited focusing) for each of the design frequencies and angles. }
    \label{fig:virfoc}
\end{figure}

Modern AR/VR applications demand increasingly sophisticated optical components with ultra-compact form-factors that can deliver the same level of performance as their conventional bulky counterparts. Metalenses and multi-level diffractive optics~\cite{yu2014flat,khorasaninejad2016metalenses,chen2018broadband,kim2012design,meem2020large,arbabi2016miniature,groever2017meta} offer an innovative solution by replacing the bulky refractive components with much more compact interfaces, but they have not fundamentally altered the familiar multi-lens stacking strategy that uses ray optics to \emph{simultaneously} minimize chromatic and off-axis aberrations (Fig.~\ref{fig:scheme}a)---a configuration which still consumes a large part of the device volume\cite{kress201711,kress2019digital}. We propose new nanophotonic solutions for miniaturization of optics, which reduce the role of ray mechanics and free-space propagation, by exploiting the much richer wave interactions occurring in nanostructured photonic media. Specifically, we present \emph{single-piece nanophotonic plan-achromats} (Fig.~\ref{fig:scheme}b) for achromatic, aplanatic and curvature-free focusing, that is, realizing diffraction-limited focal spots (Strehl ratio $> 0.8$ and absolute focusing efficiency of up to 65\%) at precisely prescribed positions on a flat sensor over finite spectral and angular bandwidths (Sections~\ref{sec:plach} and \ref{sec:axi}). Our proposed solutions also allow for shortening the lens-to-sensor distance (Fig.~\ref{fig:scheme}c) by tailoring the output angles while maintaining the plan-achromaticity (Section~\ref{sec:squeeze}). Our primary goal in this work is to demonstrate the vast number of functionalities that can be incorporated into a small thickness ($\sim 10\lambda$, where $\lambda$ is the operating wavelength) of low-index polymer structures by exploiting photonic inverse design~\cite{jensen2011topology,lalau2013adjoint,molesky2018inverse}, enabling over four orders of magnitude ($>10^4$) reduction in device thickness compared to traditional multi-lens systems. These results motivate future work to combine advances in 3D fabrication~\cite{zhan2019controlling, christiansen2020fullwave, moughames2020three, gehring2019low} with computational methods for imposing fabrication constraints~\cite{augenstein2020inverse,li2016connect,zhou2015minimum} in order to tailor these capabilities and explore manufacturing tradeoffs for specific applications. 

Prior work on metalens optics has been primarily devoted to achieving achromatic focusing at the normal incidence~\cite{chen2018broadband,wang2018broadband,shrestha2018broadband} or correcting off-axis aberrations at a single frequency~\cite{arbabi2016miniature,groever2017meta}. In particular, a number of inverse design techniques have been explored to maximize metalens efficiencies over \emph{either} spectral or angular bandwidths; such techniques involve combining locally periodic approximations, domain decomposition methods or even full-wave Maxwell solvers with large-scale optimization algorithms~\cite{jensen2011topology,molesky2018inverse} to discover multi-functional metalens geometries~\cite{sell2017large,lin2018topology,pestourie2018inverse,lin2019topology,lin2019overlapping,zhan2019controlling,chung2020high,christiansen2020fullwave}.
However, achieving single-piece plan-achromaticity (i.e. aberration-free focusing over \emph{simultaneous} spectral and angular bandwidths) remains an open problem. While plan-achromaticity may be achieved by a multiplet metalens configuration~\cite{shrestha2019multi}, such an approach still requires considerable propagation distance between individual lenses, which presents a significant hurdle to further reducing the device footprint. In contrast, our computational results indicate that it is possible to design \emph{single-piece uni-body plan-achromats}, enabled by utilizing 3D nanophotonics, thus circumventing the necessity of multiplet ray-optics designs.

\section{Inverse design of a single-piece plan-achromat}
\label{sec:plach}

In this work we employ a topology-optimization~\cite{bendsoe2013topology} based tool to design the metalenses. Topology optimization (TO) is a computational technique for inverse design that can handle extensive design spaces ( $>10^9$ design variables (DVs)~\cite{aage2017giga}), considering the dielectric permittivity at every spatial point as a DV~\cite{jensen2011topology,lalau2013adjoint,molesky2018inverse}. In contrast to heuristic search routines such as genetic algorithms~\cite{mitchell1998introduction} or particle-swarm methods~\cite{kennedy2011particle}, TO employs gradient-based optimization techniques to explore hundreds to billions of \emph{continuous} DVs. Such a capability is made possible by a rapid computation of gradients (with respect to all the DVs) via adjoint methods~\cite{strang2007computational}, which require just one additional solution of Maxwell's equations~\cite{jensen2011topology,molesky2018inverse} in each design iteration. In fact, these techniques have been gaining traction lately in the field of photonic integrated circuits, especially for the design of compact modal multiplexers and converters~\cite{piggott2015inverse}. More recently, TO-based inverse design methods have been extended to metasurfaces~\cite{sell2017large,lin2018topology,lin2019topology,lin2019overlapping,christiansen2020fullwave}. 

Here, we employ TO to design a single-piece nano-photonic plan-achromat, which focuses incoming plane waves at multiple frequencies and multiple angles. The focusing quality is measured by the Strehl ratio (SR); SR above 0.8 is typically accepted as a gold standard for aberration-free imaging~\cite{khorasaninejad2016metalenses}. SR is defined as:
\begin{align}
    \text{SR} = {P_\text{peak}\over P_\text{airy}}
\end{align}
where $P_\text{peak}$ is the peak intensity at focal point and $P_\text{airy}$ is the peak intensity of a diffraction-limited ideal airy disc normalized by the same transmitted power as that above the metalens. In particular, to design a plan-achromat, we maximize the minimum of SR's for multiple frequencies and angles $\{(\omega_i,\theta_i): i = 1,...,N\}$; such a problem can be equivalently formulated in an epigraph form to obtain a differentiable optimization problem~\cite{boyd2004convex}, using an extra variable $t$:
\begin{align}
    \max_{t,\varepsilon} \quad & t \\
    t & \le \text{SR}_i(\varepsilon),~i=1,...,N
\end{align}
In contrast to typical TO, where the relative permitivitty $\varepsilon(\mathbf{r})$ is varied at each pixel, we consider varying the thickness profile of each layer~\cite{christiansen2020fullwave} with an aim to leverage the capabilities of 3D printing~\cite{gissibl2016two,christiansen2020fullwave} and thermal scanning probe lithography~\cite{lassaline2020optical} for fabrication and characterization in future works.

In Fig.~\ref{fig:2d1}, we present a proof-of-concept 2D cylindrical lens design (invariant along $y$ axis), which can focus 10 frequencies and 10 angles ($N=100$) within a spectral and angular bandwidth of 23\% and 60$^\circ$, respectively. The lens is $50\lambda$ wide and has a numerical aperture (NA) of $0.24$; it consists of 20 layers of 3D-printable polymer (refractive index $\sim 1.5$), corresponding to a total thickness of $12\lambda$. Remarkably, TO discovers a design with a fairly uniform $SR\approx 0.89$, achieving aberration-free diffraction-limited focusing for all the design frequencies and angles; the uniformity is a typical by-product of the epigraph formulation~\cite{boyd2004convex}. (In between the frequencies and angles targeted in the design process, SR averages to above 0.7) In addition to SR, it is also instructive to compute the absolute focusing effciency (AE), defined as the fraction of transmitted power within three full-widths at half-maximum (FWHM) around the focal peak divided by the total incident power. Fig.~\ref{fig:2d1} shows simulated AE's whose average is $55\%$; while the AE of our design may fall short of the efficiencies typically seen in bulky refractive lenses, it is better than or comparable to any previous achromatic metalens design~\cite{chen2018broadband,chung2020high}, despite the mulitple scatterings and complex interactions occuring within a multi-layer geometry; further improvements in AE are likely possible if we also consider transmitted power as an additional constraint during optimization. Next, we demonstrate that our technique is not limited to small NA lenses; indeed, it can be straightforwardly scaled to higher NAs and larger lens widths. Fig.~\ref{fig:2d2} shows a wider and thicker design ($125\lambda$ wide and $24\lambda$ thick) with a substantially higher NA of 0.7, achieving $\text{SR}>0.85$ for 4 frequencies and 4 angles with an average efficiency of $25\%$. 

\section{Axisymmetric inverse design for off-axis incidence}
\label{sec:axi}

Having shown proof-of-concept 2D cylindrical lens designs, we now demonstrate that the proposed optimization framework can be used to design 3D plan-achromats. Designing a large-area fully freeform 3D plan-achromat would however demand much more intensive computational resources than were available to us in this work. A way to dramatically reduce the computational complexity is to exploit the axial symmetry (axi-symmetry) of focusing~\cite{christiansen2020fullwave}. In other words, we restrict our 3D designs to be rotationally symmetric around a central axis. Doing this allows us to describe space using polar coordinates $(r,\phi,z)$ and exploit the $\phi$-invariance $\varepsilon(r,\phi,z)=\varepsilon(r,z)$ (Fig.~\ref{fig:3d} top). This allows for the reduction of the 3D Maxwell equation to a set of independent 2D equations in $(r,z)$, each characterized by an integer angular momentum number $m$~\cite{taflove2005computational}:
\begin{align}
    {\cal D}^2 \mathbf{E}_m - \w^2 \e(r,z) \mathbf{E}_m = i \w \mathbf{J}_m,
\end{align}
where
\begin{align}
{\cal D} &= 
\begin{pmatrix}
0 		& 	-\ppz 		& 	{im \over r} \\
 \ppz	&	0			&	- \ppr \\
-{i m \over r}	&	{1 \over r}{\partial \over \partial r} r	&	0 \notag
\end{pmatrix}.
\end{align}
The total electric field $\mathbf{E}$ is then given by the \emph{coherent sum} of the individual fields times the exponential azimuthal factor ($\mathbf{E}_m e^{i m \phi}$) over all $m$'s. 

It is important to recognize that, in contrast to the axisymmetric material distribution $\varepsilon$, the incident current $\mathbf{J}$ may not necessarily obey axi-symmetry. For example, linearly-polarized normally-incident plane waves may be decomposed into two counter-rotating circularly-polarized waves ($m=\pm 1$) both of which do obey axi-symmetry. In contrast, any plane wave impinging on the lens at an oblique angle $\theta$ must be decomposed into an infinite number of axisymmetric components (with different $m$'s), the sum of which may be truncated at a finite number of terms, depending on the desired accuracy of the approximation; this decomposition (known as the Jacobi--Anger expansion) is given by:
\begin{align}
    \mathbf{J} &= A_x e^{ik_x x + ik_y y} \xhat + A_y e^{ik_x x + ik_y y} \yhat \\
    &= \sum_m J_r^m e^{im\phi} \rhat + \sum_m J_\phi^m e^{im\phi} \mathbf{\hat{\phi}} \\
    \sqrt{ k_x^2 + k_y^2 } &= {2 \pi \over \lambda} \sin{\theta}, \\
     J_r^m   &=  A_x P_m + A_y Q_m, \quad 
    J_\phi^m =  A_y P_m - A_x Q_m, \\
    P_m(r) &= \int_0^{2\pi} \exp\Big[ ik_x r \cos{\phi} + ik_y r \sin{\phi} - i m \phi \Big] \cos{\phi}  ~d\phi \\
    Q_m(r) &= \int_0^{2\pi} \exp\Big[ ik_x r \cos{\phi} + ik_y r \sin{\phi} - i m \phi \Big] \sin{\phi}  ~d\phi.
\end{align}
where the integrals $P$ and $Q$ can be evaluated by Bessel's functions. Therefore, the total electric field response of an axisymmetric lens under off-axis plane-wave incidence is a coherent sum of individual fields in response to the current components $\mathbf{J}_m$, which can be computed in a highly parallel fashion.  

We combine the axisymmetric Maxwell solver with topology optimization to design a fully-3D single-piece plan-achromat (Fig.~\ref{fig:3d}), rigorously taking into account off-axis angular aberrations. The lens has a diameter of $50\lambda$ and a thickness of $12\lambda$ with an NA of 0.12. It is optimized for aberration-free focusing at 5 frequencies and 5 angles ($N=25$) within 23\% bandwidth and 16$^\circ$ field of view. The lens achieves diffraction-limited focusing with SR > 0.86 for all the design frequencies and angles while the average absolute efficiency is 72\%. To accurately model the off-axis angles, we consider up to 50 angular mode numbers, distributing our simulations over $\sim 1000$ CPUs with a computational time of about a week. Much larger diameters and numerical apertures can be designed by combining our approach with an overlapping domains method~\cite{lin2019overlapping}.

\section{Plan-achromat with a virtual focal length}
\label{sec:squeeze}

While we have shown that it is possible to dramatically reduce a cm-scale multi-plet plan-achromat to a micron-scale single-piece with nano-scale 3D structuring, the overall system may still remain undesirably bulky because of the free space between the lens and the sensor~\cite{reshef2020towards}. We note that to further reduce this distance is not a simple matter of just designing a lens with a shorter focal length, because conventional ray optics dictates that in order to achieve a desired magnification at a finite sensor resolution, the lens and sensor must be sufficiently separated. Conventionally, ``compact imitations'' such as a telephoto array are used to mimic the effect of a long-focus lens within a physically shorter device. On the other hand, it should be clear that the actual distance between the optical component and the sensor should not matter as long as the sensor measures the same output for any given input in the compact imitation as in the long-focus original. Since an arbitrary input can be decomposed into a sum of plane waves, the compact imitation must reproduce, for each input angle $\theta$, a corresponding focal spot at the same position on a near sensor as the original would produce on a far sensor, ie, the focal spot for $\theta$ must be formed at $f\tan{\theta}$ away from the sensor center where $f$ is the virtual focal length of a desired long-focus lens, which may be longer than the actual lens-sensor distance $f' > f$ in the compact equivalent (Fig.~\ref{fig:virfoc}-top left inset and also Fig.~\ref{fig:scheme}c). This effectively means that any nano-structured equivalent, which labors to shorten the lens-sensor distance, must stretch the output angle as specified by the relation $\theta' = \tan^{-1} \lp {f \over f'} \tan{\theta} \rp$. Our inverse design formulation is naturally suited to design for such effects because we can explicitly specify an arbitrary input-output relation to our device~\cite{christiansen2016negreffulldomain}. Here, we note that recently, an alternative approach to squeezing free space has also been proposed based on non-local metamaterials~\cite{reshef2020towards,guo2020squeeze}; in contrast, our approach specifically considers the space-squeezing problem in an integral fashion, simultaneously with the focusing problem and seeks to discover an ``all-in-one uni-body'' nanophotonic space-squeezing plan-achromat. 

Fig.~\ref{fig:virfoc} shows a proof-of-concept 2D plan-achromat space-squeezer with a \emph{virtual} focal length of $200\lambda$ but a physical lens-to-sensor distance of $130\lambda$ (54\% compression). The lens is $50\lambda$ wide and $12\lambda$ thick, with a \emph{virtual} NA of 0.12, and is optimized to perform at 10 frequencies and 10 angles over spectral and angular bandwidths of $23\%$ and $16^\circ$ respectively, achieving \emph{virtual} SR > 0.8 for all the design frequencies and angles (Note that the \emph{virtual} SR is defined relative to the \emph{virtual} airy disc, i.e., the ideal airy disc that would have been formed at the \emph{virtual} focal distance of $200\lambda$). One trade-off we found in the space-squeezer design is that the bigger the compression, the smaller the absolute focusing efficiency even for a small numerical aperture. While additional optimization constraints may be used to improve the transmission efficiency, we speculate that there exists a fundamental trade-off between compression and transmission---a theoretical question we will seek to elucidate in future work. On the other hand, such a trade-off may be entirely circumvented in an end-to-end framework, in which a front-end photonic design is fully coupled with a back-end computational reconstruction algorithm~\cite{lin2020end}.  

\section{Summary and outlook}

In summary, we have presented computational designs for single-piece plan achromats with SR > 0.8 and focusing efficiencies up to 71\%. We have also demonstrated that by solving the Maxwell equations under axial symmetry, one can design 3D axisymmetric plan-achromats that fully consider off-axis propagation. Our solutions also suggest the possibility of further reducing the lens-to-sensor distance by manipulating the relation between input and output angles and mimicking a longer \emph{virtual} focal length. 

Our optimizations were performed over $\sim 1000$ CPUs within one day to a week. We surmise that it is possible to design high-NA single-piece plan-achromats with millimeters in diameter and $\sim 100~\mathrm{\mu m}$ in thickness, given sufficient computational resources on the order of $\sim 3000$ CPUs combined with advanced numerical solvers which exploits the overlapping domains technique~\cite{lin2019overlapping}.
 
Our theoretical results demonstrate the untapped potential of volumetric nanophotonics and sets the foundations for future works, which will be devoted to developing 3D devices tailored for specific fabrication technologies.  For any given fabrication technology, there are a wealth of computational techniques to impose corresponding design constraints on feature sizes~\cite{zhou2015minimum}, connectivity~\cite{li2016connect}, and mechanical stability~\cite{augenstein2020inverse}. For our proof-of-concept designs in this paper, it was convenient to parameterize the structures in terms of variable-height layers, leading to disconnected multi-layer designs, but this is merely an artifact of this parameterization and is not fundamental to plan-achromat functionality.  We do not expect that performance will degrade for connected structures designed by density-based topology optimization (where every pixel is a degree of freedom rather than the layer thicknesses) with connectivity constraints.
 
Experimentally, there has been a surge of interest in fabricating multilayer meta-optics devices~\cite{mansouree2020multifunctional, zhou2018multilayer}. These designs require fabrication techniques more advanced than traditional CMOS single-step lithography, such as multi-step or multi-photon lithography. While these techniques have been primarily employed for micro-scale 3D fabrication~\cite{gissibl2016two}, recent work has demonstrated their potential in 3D nanophotonic interconnects~\cite{moughames2020three}, fiber couplers~\cite{gehring2019low}, and meta-optics~\cite{zhan2019controlling,christiansen2020fullwave}, suggesting that two-photon lithography could enable a whole new range of applications in 3D nanophotonics. Our work motivates further efforts to take advantage of emerging advances in 3D fabrication by demonstrating the potential of inverse-designed 3D meta-optics.

We note that our results represent an all-optical solution to imaging that does not require extensive calibration measurements. An alternative strategy that can ease the burden on optics is to measure ``imperfect'' point-spread functions and apply image processing algorithms to extract a faithful image~\cite{tarantola2005inverse}. Recently, we have also shown that nanophotonic structures can be inverse-designed in a holistic end-to-end fashion together with any image processing backend~\cite{lin2020end}. Ultimately, we expect that an optimal mix of hardware and software can be found to solve the most challenging problems in modern optical technologies as well as to pave the way for extraordinary functionalities that have not yet been explored.

\section*{Funding} 
This work was supported in part by Villum Fonden through the NATEC (NAnophotonics for TErabit Communications) Centre (grant no.~8692); the Danish National Research Foundation through NanoPhoton Center for Nanophotonics (grant no.~DNRF147); the U.~S.~Army Research Office through the Institute for Soldier Nanotechnologies (award no.~W911NF-18-2-0048); and the MIT-IBM Watson AI Laboratory (challenge no.~2415).

%\bibliographystyle{unsrt}
%\bibliography{ref}

\begin{thebibliography}{10}

\bibitem{yu2014flat}
Nanfang Yu and Federico Capasso.
\newblock Flat optics with designer metasurfaces.
\newblock {\em Nature Materials}, 13(2):139, 2014.

\bibitem{khorasaninejad2016metalenses}
Mohammadreza Khorasaninejad, Wei~Ting Chen, Robert~C Devlin, Jaewon Oh,
  Alexander~Y Zhu, and Federico Capasso.
\newblock Metalenses at visible wavelengths: Diffraction-limited focusing and
  subwavelength resolution imaging.
\newblock {\em {Science}}, 352(6290):1190--1194, 2016.

\bibitem{chen2018broadband}
Wei~Ting Chen, Alexander~Y Zhu, Vyshakh Sanjeev, Mohammadreza Khorasaninejad,
  Zhujun Shi, Eric Lee, and Federico Capasso.
\newblock A broadband achromatic metalens for focusing and imaging in the
  visible.
\newblock {\em Nature Nanotechnology}, 13(3):220, 2018.

\bibitem{kim2012design}
Ganghun Kim, Jos{\'e}~A Dom{\'\i}nguez-Caballero, and Rajesh Menon.
\newblock Design and analysis of multi-wavelength diffractive optics.
\newblock {\em {Optics Express}}, 20(3):2814--2823, 2012.

\bibitem{meem2020large}
Monjurul Meem, Sourangsu Banerji, Christian Pies, Timo Oberbiermann, Apratim
  Majumder, Berardi Sensale-Rodriguez, and Rajesh Menon.
\newblock Large-area, high-numerical-aperture multi-level diffractive lens via
  inverse design.
\newblock {\em Optica}, 7(3):252--253, 2020.

\bibitem{arbabi2016miniature}
Amir Arbabi, Ehsan Arbabi, Seyedeh~Mahsa Kamali, Yu~Horie, Seunghoon Han, and
  Andrei Faraon.
\newblock Miniature optical planar camera based on a wide-angle metasurface
  doublet corrected for monochromatic aberrations.
\newblock {\em Nature communications}, 7(1):1--9, 2016.

\bibitem{groever2017meta}
Benedikt Groever, Wei~Ting Chen, and Federico Capasso.
\newblock Meta-lens doublet in the visible region.
\newblock {\em Nano letters}, 17(8):4902--4907, 2017.

\bibitem{kress201711}
Bernard~C Kress and William~J Cummings.
\newblock 11-1: Invited paper: Towards the ultimate mixed reality experience:
  Hololens display architecture choices.
\newblock In {\em SID Symposium Digest of Technical Papers}, volume~48, pages
  127--131. Wiley Online Library, 2017.

\bibitem{kress2019digital}
Bernard~C Kress.
\newblock Digital optical elements and technologies (edo19): applications to
  ar/vr/mr.
\newblock In {\em Digital Optical Technologies 2019}, volume 11062, page
  1106222. International Society for Optics and Photonics, 2019.

\bibitem{jensen2011topology}
Jakob~S{\o}ndergaard Jensen and Ole Sigmund.
\newblock Topology optimization for nano-photonics.
\newblock {\em Laser \& Photonics Reviews}, 5(2):308--321, 2011.

\bibitem{lalau2013adjoint}
Christopher~M Lalau-Keraly, Samarth Bhargava, Owen~D Miller, and Eli
  Yablonovitch.
\newblock Adjoint shape optimization applied to electromagnetic design.
\newblock {\em Optics Express}, 21(18):21693--21701, 2013.

\bibitem{molesky2018inverse}
Sean Molesky, Zin Lin, Alexander~Y Piggott, Weiliang Jin, Jelena Vuckovi{\'c},
  and Alejandro~W Rodriguez.
\newblock Inverse design in nanophotonics.
\newblock {\em Nature Photonics}, 12(11):659, 2018.

\bibitem{zhan2019controlling}
Alan Zhan, Ricky Gibson, James Whitehead, Evan Smith, Joshua~R Hendrickson, and
  Arka Majumdar.
\newblock Controlling three-dimensional optical fields via inverse mie
  scattering.
\newblock {\em Science Advances}, 5(10):eaax4769, 2019.

\bibitem{christiansen2020fullwave}
Rasmus~E Christiansen, Zin Lin, Charles~Roques Carmes, Yannick Salamin,
  Steven~E Kooi, John~D Joannopoulos, Marin Solja{\v{c}}i{\'c}, and Steven~G
  Johnson.
\newblock Fullwave maxwell inverse design of axisymmetric, tunable, and
  multi-scale multi-wavelength metalenses.
\newblock {\em Optics Express}, 2020.
\newblock In press, arXiv preprint arXiv:2007.11661.

\bibitem{moughames2020three}
Johnny Moughames, Xavier Porte, Michael Thiel, Gwenn Ulliac, Laurent Larger,
  Maxime Jacquot, Muamer Kadic, and Daniel Brunner.
\newblock Three-dimensional waveguide interconnects for scalable integration of
  photonic neural networks.
\newblock {\em Optica}, 7(6):640--646, 2020.

\bibitem{gehring2019low}
H~Gehring, M~Blaicher, W~Hartmann, P~Varytis, K~Busch, M~Wegener, and WHP
  Pernice.
\newblock Low-loss fiber-to-chip couplers with ultrawide optical bandwidth.
\newblock {\em APL Photonics}, 4(1):010801, 2019.

\bibitem{augenstein2020inverse}
Yannick Augenstein and Carsten Rockstuhl.
\newblock Inverse design of nanophotonic devices with structural integrity.
\newblock {\em ACS photonics}, 7(8):2190--2196, 2020.

\bibitem{li2016connect}
Quhao Li, Chen Wenjiong, Liu Shutian, and Liyong Tong.
\newblock Structural topology optimization considering connectivity constraint.
\newblock {\em Struct Multidisc Optim}, 54:971--984, 2016.

\bibitem{zhou2015minimum}
Mingdong Zhou, Boyan~S Lazarov, Fengwen Wang, and Ole Sigmund.
\newblock Minimum length scale in topology optimization by geometric
  constraints.
\newblock {\em Computer Methods in Applied Mechanics and Engineering},
  293:266--282, 2015.

\bibitem{wang2018broadband}
Shuming Wang, Pin~Chieh Wu, Vin-Cent Su, Yi-Chieh Lai, Mu-Ku Chen, Hsin~Yu Kuo,
  Bo~Han Chen, Yu~Han Chen, Tzu-Ting Huang, Jung-Hsi Wang, et~al.
\newblock A broadband achromatic metalens in the visible.
\newblock {\em Nature Nanotechnology}, 13(3):227, 2018.

\bibitem{shrestha2018broadband}
Sajan Shrestha, Adam~C Overvig, Ming Lu, Aaron Stein, and Nanfang Yu.
\newblock Broadband achromatic dielectric metalenses.
\newblock {\em Light: Science \& Applications}, 7(1):85, 2018.

\bibitem{sell2017large}
David Sell, Jianji Yang, Sage Doshay, Rui Yang, and Jonathan~A Fan.
\newblock Large-angle, multifunctional metagratings based on freeform multimode
  geometries.
\newblock {\em {Nano Letters}}, 17(6):3752--3757, 2017.

\bibitem{lin2018topology}
Zin Lin, Benedikt Groever, Federico Capasso, Alejandro~W Rodriguez, and Marko
  Lon{\v{c}}ar.
\newblock Topology-optimized multilayered metaoptics.
\newblock {\em Physical Review Applied}, 9(4):044030, 2018.

\bibitem{pestourie2018inverse}
Rapha{\"e}l Pestourie, Carlos P{\'e}rez-Arancibia, Zin Lin, Wonseok Shin,
  Federico Capasso, and Steven~G Johnson.
\newblock Inverse design of large-area metasurfaces.
\newblock {\em {Optics Express}}, 26(26):33732--33747, 2018.

\bibitem{lin2019topology}
Zin Lin, Victor Liu, Rapha{\"e}l Pestourie, and Steven~G Johnson.
\newblock Topology optimization of freeform large-area metasurfaces.
\newblock {\em Optics Express}, 27(11):15765--15775, 2019.

\bibitem{lin2019overlapping}
Zin Lin and Steven~G Johnson.
\newblock Overlapping domains for topology optimization of large-area
  metasurfaces.
\newblock {\em Optics Express}, 27(22):32445--32453, 2019.

\bibitem{chung2020high}
Haejun Chung and Owen~D Miller.
\newblock High-{NA} achromatic metalenses by inverse design.
\newblock {\em Optics Express}, 28(5):6945--6965, 2020.

\bibitem{shrestha2019multi}
Sajan Shrestha, Adam Overvig, and Nanfang Yu.
\newblock Multi-element meta-lens systems for imaging.
\newblock In {\em 2019 Conference on Lasers and Electro-Optics (CLEO)}, pages
  1--2. IEEE, 2019.

\bibitem{bendsoe2013topology}
Martin~Philip Bendsoe and Ole Sigmund.
\newblock {\em Topology optimization: theory, methods, and applications}.
\newblock Springer Science \& Business Media, 2013.

\bibitem{aage2017giga}
Niels Aage, Erik Andreassen, Boyan~S Lazarov, and Ole Sigmund.
\newblock Giga-voxel computational morphogenesis for structural design.
\newblock {\em Nature}, 550(7674):84--86, 2017.

\bibitem{mitchell1998introduction}
Melanie Mitchell.
\newblock {\em {An Introduction to Genetic Algorithms}}.
\newblock MIT press, 1998.

\bibitem{kennedy2011particle}
James Kennedy.
\newblock Particle swarm optimization.
\newblock In {\em Encyclopedia of Machine Learning}, pages 760--766. Springer,
  2011.

\bibitem{strang2007computational}
Gilbert Strang.
\newblock {\em {Computational Science and Engineering}}, volume 791.
\newblock Wellesley-Cambridge Press Wellesley, 2007.

\bibitem{piggott2015inverse}
Alexander~Y Piggott, Jesse Lu, Konstantinos~G Lagoudakis, Jan Petykiewicz,
  Thomas~M Babinec, and Jelena Vu{\v{c}}kovi{\'c}.
\newblock Inverse design and demonstration of a compact and broadband on-chip
  wavelength demultiplexer.
\newblock {\em Nature Photonics}, 9(6):374--377, 2015.

\bibitem{boyd2004convex}
Stephen Boyd, Stephen~P Boyd, and Lieven Vandenberghe.
\newblock {\em {Convex Optimization}}.
\newblock Cambridge University Press, 2004.

\bibitem{gissibl2016two}
Timo Gissibl, Simon Thiele, Alois Herkommer, and Harald Giessen.
\newblock Two-photon direct laser writing of ultracompact multi-lens
  objectives.
\newblock {\em Nature Photonics}, 10(8):554--560, 2016.

\bibitem{lassaline2020optical}
Nolan Lassaline, Raphael Brechb{\"u}hler, Sander~JW Vonk, Korneel Ridderbeek,
  Martin Spieser, Samuel Bisig, Boris le~Feber, Freddy~T Rabouw, and David~J
  Norris.
\newblock Optical fourier surfaces.
\newblock {\em Nature}, 582(7813):506--510, 2020.

\bibitem{taflove2005computational}
Allen Taflove and Susan~C Hagness.
\newblock {\em {Computational Electrodynamics: the Finite-Difference
  Time-Domain Method}}.
\newblock Artech house, 2005.

\bibitem{reshef2020towards}
Orad Reshef, Michael~P DelMastro, Katherine~KM Bearne, Ali~H Alhulaymi, Lambert
  Giner, Robert~W Boyd, and Jeff~S Lundeen.
\newblock Towards ultra-thin imaging systems: an optic that replaces space.
\newblock {\em arXiv preprint arXiv:2002.06791}, 2020.

\bibitem{christiansen2016negreffulldomain}
Rasmus~E Christiansen and Ole Sigmund.
\newblock Designing meta material slabs exhibiting negative refraction using
  topology optimization.
\newblock {\em Struct Multidisc Optim}, 54:469--482, 2016.

\bibitem{guo2020squeeze}
Cheng Guo, Haiwen Wang, and Shanhui Fan.
\newblock Squeeze free space with nonlocal flat optics.
\newblock {\em arXiv preprint arXiv:2003.06918}, 2020.

\bibitem{lin2020end}
Zin Lin, Charles Roques-Carmes, Rapha{\"e}l Pestourie, Marin
  Solja{\v{c}}i{\'c}, Arka Majumdar, and Steven~G Johnson.
\newblock End-to-end inverse design for inverse scattering via freeform
  metastructures.
\newblock {\em arXiv preprint arXiv:2006.09145}, 2020.

\bibitem{mansouree2020multifunctional}
Mahdad Mansouree, Hyounghan Kwon, Ehsan Arbabi, Andrew McClung, Andrei Faraon,
  and Amir Arbabi.
\newblock Multifunctional 2.5 d metastructures enabled by adjoint optimization.
\newblock {\em Optica}, 7(1):77--84, 2020.

\bibitem{zhou2018multilayer}
You Zhou, Ivan~I Kravchenko, Hao Wang, J~Ryan Nolen, Gong Gu, and Jason
  Valentine.
\newblock Multilayer noninteracting dielectric metasurfaces for multiwavelength
  metaoptics.
\newblock {\em Nano letters}, 18(12):7529--7537, 2018.

\bibitem{tarantola2005inverse}
Albert Tarantola.
\newblock {\em {Inverse Problem Theory and Methods for Model Parameter
  Estimation}}.
\newblock SIAM, 2005.

\end{thebibliography}

\end{document}